\documentclass[preprint,nopreprintline]{elsarticle}

\usepackage{siunitx}
\mathchardef\mhyphen="2D
\DeclareSIUnit\e{e{\ensuremath{^\mathrm{\mhyphen}}}}
\DeclareSIUnit\rad{rad}
\usepackage{booktabs}

\usepackage{graphicx}
\usepackage{amsmath}
\usepackage{multirow}

\begin{document}

\begin{frontmatter}

\title{X-ray Irradiation Studies on the Monopix DMAPS in 150$\,$nm and 180$\,$nm}

\author[1]{Christian Bespin\corref{cor1}}%
\ead{cbespin@uni-bonn.de}
\author[2]{Marlon Barbero}
\author[2]{Pierre Barrillon}
\author[2]{Patrick Breugnon}
\author[1]{Ivan Caicedo}
\author[3]{Yavuz Degerli}
\author[1]{Jochen Dingfelder}
\author[1]{Tomasz Hemperek\fnref{fn1}}
\author[1]{Toko Hirono\fnref{f2}}
\author[1]{Hans Kr\"uger}
\author[1]{Fabian H\"ugging}
\author[1]{Konstantinos Moustakas\fnref{fn3}}
\author[2]{Patrick Pangaud}
\author[4]{Heinz Pernegger}
\author[4]{Petra Riedler}
\author[1]{Piotr Rymaszewski\fnref{fn1}}
\author[1]{Lars Schall\corref{cor1}}
\author[3]{Philippe Schwemling}
\author[4]{Walter Snoeys}
\author[1]{Tianyang Wang\fnref{4}}
\author[1]{Norbert Wermes}
\author[1]{Sinou Zhang}

\cortext[cor1]{Corresponding author}
\fntext[fn1]{now at DECTRIS AG, Baden-D\"attwil, Switzerland}
\fntext[fn2]{now at Karlsruher Institut für Technologie, Karlsruhe, Germany}
\fntext[fn3]{now at Paul Scherrer Institut, Villingen, Switzerland}
\fntext[fn4]{now at Zhangjiang National Laboratory, Pudong, China}

\affiliation[1]{organization={Physikalisches Institut der Universität Bonn}, 
                 addressline={Nu{\ss}allee 12},
                 postcode={53115}, 
                 city={Bonn}, 
                 country={Germany}}
\affiliation[2]{organization={CPPM, Aix Marseille Universit\'e}, 
                 addressline={163 Avenue de Luminy},
                 postcode={13288}, 
                 city={Marseille}, 
                 country={France}}
\affiliation[3]{organization={IRFU CEA/Saclay}, 
                 addressline={B\^atiment 141},
                 postcode={91191}, 
                 city={Gif-sur-Yvette}, 
                 country={France}}
\affiliation[4]{organization={CERN}, 
                 addressline={Espl. des Particules 1},
                 postcode={1217}, 
                 city={Geneva}, 
                 country={Switzerland}}

\begin{abstract}
    Monolithic active pixel sensors with depleted substrates present a promising option for pixel detectors in high-radiation environments.
    High-resistivity silicon substrates and high bias voltage capabilities in commercial CMOS technologies facilitate depletion of the charge sensitive volume.
    TJ-Monopix2 and LF-Monopix2 are the most recent large-scale chips in their respective development line, aiming for the ATLAS Inner Tracker outer layer requirements.
    Those include a tolerance to ionizing radiation of up to~\qty{100}{\mega\rad}.
    It was evaluated by irradiating both devices with X-rays to the corresponding ionization dose, showing no significant degradation of the performance at~\qty{100}{\mega\rad} and continuous operability throughout the irradiation campaign.
\end{abstract}


\end{frontmatter}

\section{Introduction}
The fabrication of monolithic pixel detectors for high-energy physics experiments in commercial CMOS processes has been successfully demonstrated in many projects and chips in recent years.
With the original design specifications to meet the ATLAS Inner Tracker outer pixel layer requirements, the development of LF- and TJ-Monopix was heavily influenced by the expected hit rate and radiation levels in terms of ionizing and non-ionizing radiation.

\section{The Monopix2 Chips}
Both monolithic pixel detector prototypes, LF- and TJ-Monopix2, utilize commercial CMOS imaging technologies with~\qty{150}{\nano\meter} and~\qty{180}{\nano\meter} feature size, respectively.
They are fabricated in highly resistive silicon and employ technologies with high-voltage capabilities to enhance depletion of the charge-sensitive sensor substrate.
The pixels are read out using a column-drain mechanism derived from the FE-I3 ATLAS readout chip~\cite{Peric2006}.
Both chips feature time over threshold charge measurements and in-pixel threshold tuning circuitry.

\subsection{LF-Monopix2}
\begin{figure}[ht]
  \centering
  \includegraphics[width=\columnwidth]{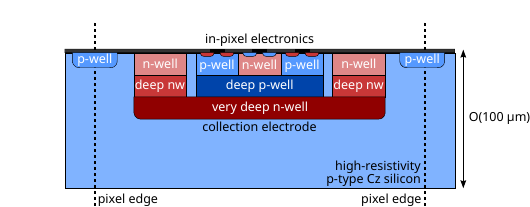}
  \caption{Schematic cross-section of a pixel in the large collection electrode design of LF-Monopix2.}
  \label{fig:lf-crosssection}
\end{figure}

The LF-Monopix2 chip features~\qtyproduct{150x50}{\micro\meter} large pixels arranged in a matrix of~\qtyproduct[product-units=single]{56x340}{pixels}.
Its large n-well collection electrode houses the analog and digital pixel electronics inside.
This design approach offers short drift distances, high and homogeneous electrical fields for fast charge collection, and high NIEL (non-ionizing energy loss) radiation tolerance~\cite{Schall2024}.
It exhibits a detector capacitance in the order of~\qty{250}{\femto\farad} which results in a power consumption of about~\qty{28}{\micro\watt\per pixel}~\cite{Caicedo2024}.
Figure~\ref{fig:lf-crosssection} shows the schematic cross-section of the pixel design.
An elaborated guard ring designs allows for bias voltages above~\qty{450}{\volt} before irradiation~\cite{Caicedo2024}.

\subsection{TJ-Monopix2}
\begin{figure}[ht]
  \centering
  \includegraphics[width=\columnwidth]{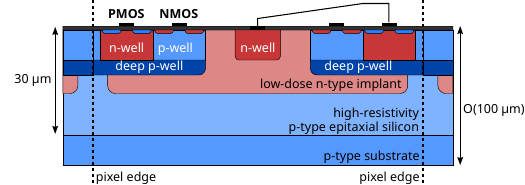}
  \caption{Schematic cross-section of a pixel in the small collection electrode design of TJ-Monopix2.}
  \label{fig:tj-crosssection}
\end{figure}

TJ-Monopix2 is a large-scale chip with a~\qtyproduct{17x17}{\milli\meter} large matrix composed of square pixels with~\qty{33}{\micro\meter} pitch.
The pixel electronics are implemented in p-wells that are spatially separated from the small (\qty{2}{\micro\meter} diameter) n-type charge collection electrode.
This design facilitates a small detector capacitance in the order of~\qty{2}{\femto\farad} enabling an analog power consumption of as low as~\qty[per-mode=symbol]{1}{\micro\watt\per pixel}~\cite{Moustakas2021}.
Due to the longer drift distances and inhomogeneous electrical field compared to the large collection electrode design, further enhancements are necessary to achieve tolerance against NIEL radiation.
A low-dose n-type layer close to the top side of the sensor leads to a depletion boundary parallel to the pixel and a more homogeneous depletion of the highly-resistive epitaxial silicon~\cite{Snoeys2017}.
The cross-section of this geometry is depicted in Figure~\ref{fig:tj-crosssection}. 
The gaps of this n-type implantation at the pixel edges improve the field shape below the p-wells in which the electronics are located~\cite{Munker2019,Dyndal2020}.

\section{X-ray Irradiation Setup and Dosimetry}

\begin{figure}[ht]
  \centering
   \includegraphics[width=.9\columnwidth]{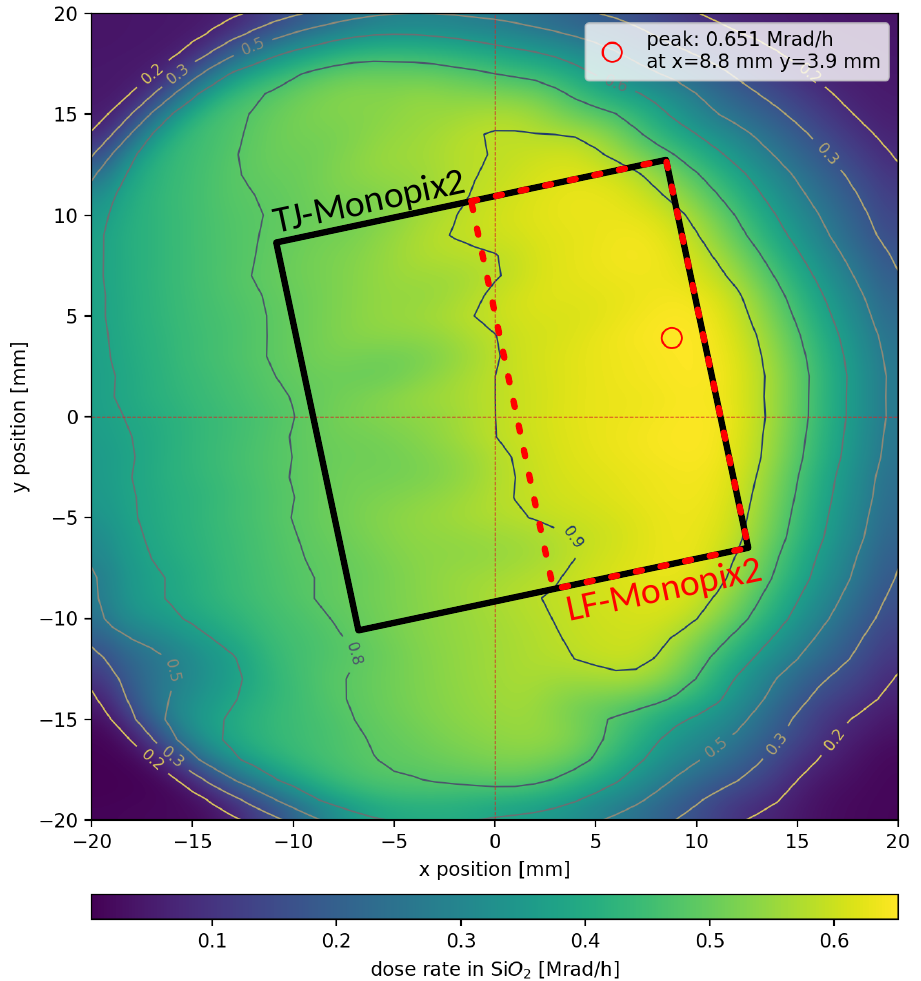}
  \caption{Dose rate map of the irradiation setup, measured with a silicon diode. The positions of the TJ- and LF-Monopix2 devices under test are depicted by the black and red rectangles, respectively.}
  \label{fig:dosimetry}
\end{figure}

The irradiation campaigns were performed at an X-ray irradiation system with a Tungsten anode at the University of Bonn~\cite{Qamesh2019}.
Figure~\ref{fig:dosimetry} depicts the dose rate profile measured with a silicon diode from which the dose rate during the irradiation is extracted.
The chips are placed in the central part to maximize the dose rate on the respective device under test.
During the irradiation, the devices under test are powered up and cooled to~\qty{0}{\degreeCelsius} by mounting them on a cold plate.

\section{Measurement Results of LF-Monopix2}
Two front-end variants of LF-Monopix2 were studied that have a similar amplifier design, but different gain.
\begin{figure}[ht]
  \centering
  \includegraphics[width=.9\columnwidth]{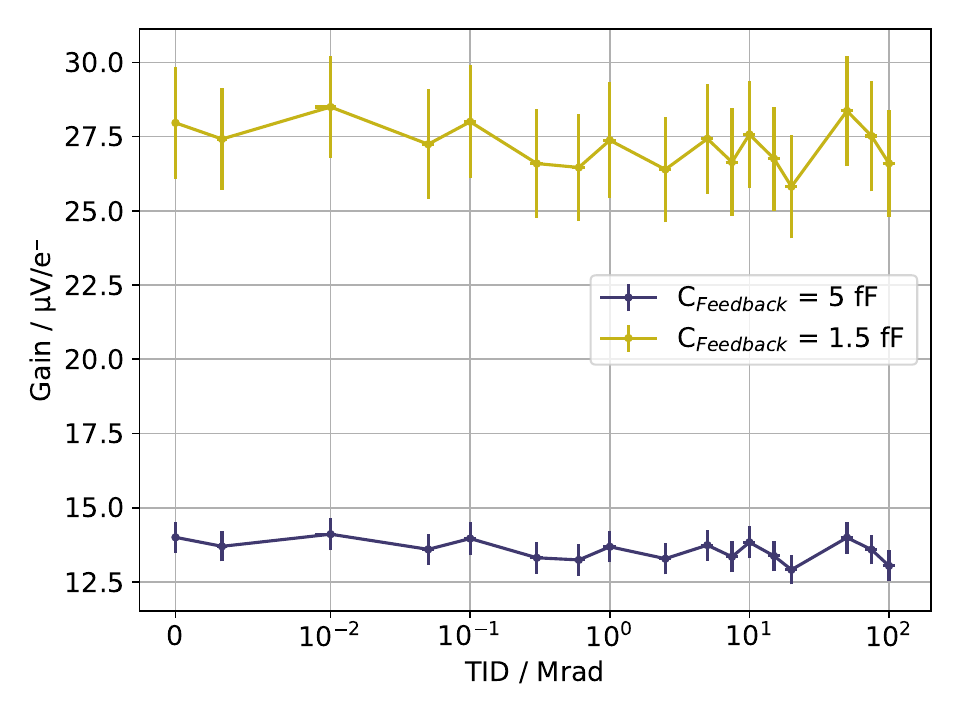}
  \caption{Evolution of the gain of LF-Monopix2 versus total ionization dose for two front-ends that differ in their feedback capacitance. For both, the gain does not degrade over the whole dose range.}
  \label{fig:lf_gain}
\end{figure}
Figure~\ref{fig:lf_gain} depicts the gain of the tested front-ends with their respective feedback capacitance against the deposited dose.
No significant change can be observed over the full dose range which is in agreement with earlier measurements on this amplifier design~\cite{Hirono2019}.
The same study finds that the feedback current in this amplifier design is also unaffected.

\begin{figure}[ht]
  \centering
  \includegraphics[width=.9\columnwidth]{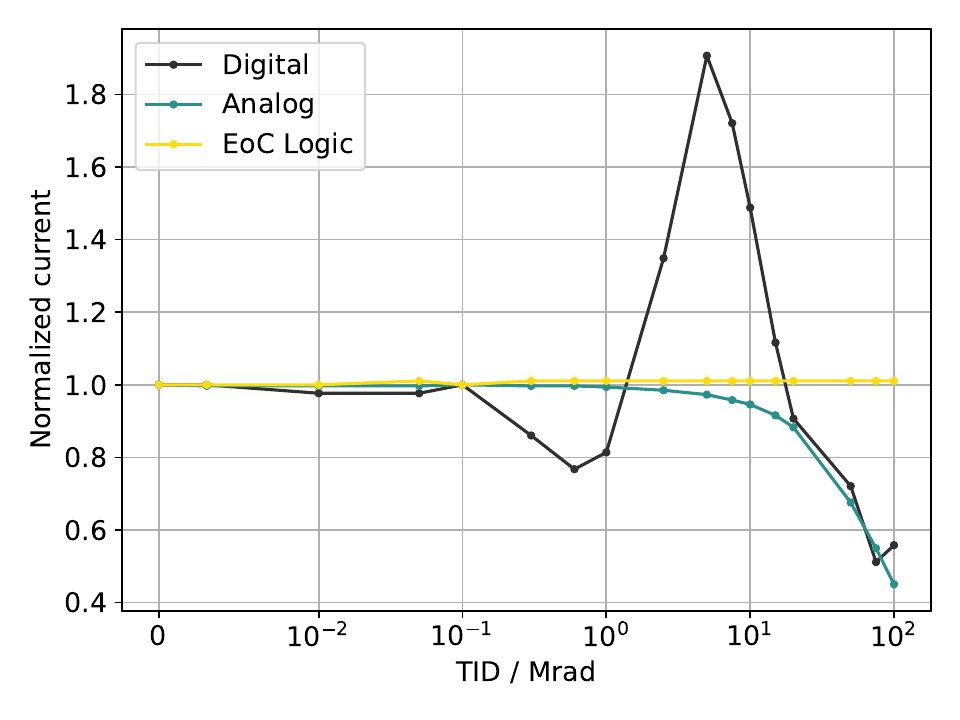}
  \caption{Normalized current consumption (at~\qty{1.8}{\volt}) of the different power domains of LF-Monopix2 versus the ionization dose.}
  \label{fig:lf_power}
\end{figure}

Monitoring the different power domains of LF-Monopix2 (for the analog and digital pixel electronics, and the end-of-column logic) enables to investigate the X-ray dose influence on the different circuitry within the chip.
The corresponding graphs are presented in Figure~\ref{fig:lf_power}.
While the current consumption of the end-of-column logic is unaffected, the current consumption of the analog circuits decreases from~\qty{1}{\mega\rad} 
on to about~\qty{50}{\percent} of its initial value.
The digital circuitry exhibits a decrease between~\qty{100}{\kilo\rad} and~\qty{1}{\mega\rad} and increases to almost twice of its initial value between~\qtylist{1;10}{\mega\rad}
It follows the behavior of the analog power domain at higher dose levels.
This increase has been observed for NMOS transistors of a comparable feature size in multiple studies.
The stated underlying reasons are two-fold~\cite{Faccio2005,Gonella2007}:
\begin{description}
  \item[Hole accumulation] Radiation\--induced positive char\-ges get trapped in the silicon oxide and introduce a threshold voltage shift that is proportional to the amount of trapped charges and depends on the distance thereof to the Si-SiO$_2$ interface.
  \item[Interface traps] For NMOS transistors, negative charges are trapped in so-called interface traps and counteract the aforementioned threshold voltage shift. 
\end{description}
Since the analog part uses well-shielded NMOS transistors, only the effect of interface traps can be observed for higher doses.
The threshold shift of the NMOS transistors in the digital circuitry increases their leakage current and, consequently, the current consumption.

An influence on the operation can be observed in the threshold dispersion that shows an increase of about a factor of four at~\qty{100}{\mega\rad} compared to the non-irradiated state.
This dispersion limits the operational threshold since pixels at the lower end of the distribution exhibit thresholds below the noise level.
The in-pixel threshold tuning in LF-Monopix2 enables compensation of the threshold variations across the matrix and limits the threshold dispersion to an increase of up to~\qty{25}{\percent} over the measured dose range, resulting in the same operational threshold of \qty{1980}{\e} at~\qty{100}{\mega\rad} compared to~\qty{2050}{\e} at~\qty{0}{\mega\rad}.

\section{Measurement Results of TJ-Monopix2}

\begin{figure}[ht]
  \centering
  \includegraphics[width=.9\columnwidth]{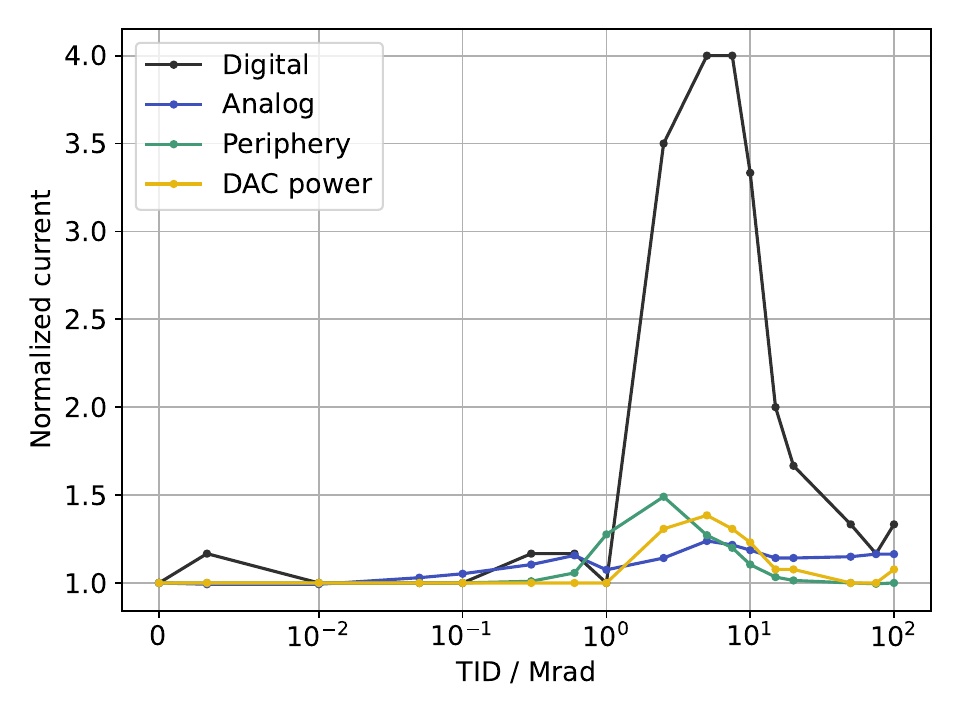}
  \caption{Normalized current consumption (at~\qty{1.8}{\volt}) of the different power domains of TJ-Monopix2 versus the ionization dose.}
  \label{fig:tj_power}
\end{figure}

The power consumption of TJ-Monopix2 versus the total irradiation dose, depicted in Figure~\ref{fig:tj_power} exhibits a similar behavior as in LF-Monopix2, originating from the same effects.
Since the voltage amplifier input transistor is implemented as an NMOS device, a significant degradation of the gain has been observed in the expected dose range between \qtylist{1;10}{\mega\rad}.
This conclusion is supported by measuring the baseline after the amplifier stage that shows the same decrease in the aforementioned dose range.
As in LF-Monopix2, the threshold dispersion increases, but can be counter-acted by utilizing the in-pixel threshold tuning capabilities.
This facilitates an operational threshold of~\qty{245}{\e} after~\qty{100}{\mega\rad} which is an increase of only~\qty{15}{\e} of the threshold in the non-irradiated chip.

\subsection{Beam Tests}

Furthermore, the irradiated sensor has been measured in a beam test campaign at the DESY II test beam facility~\cite{Diener2019}.
The device has been cooled to~\qty{0}{\degreeCelsius} as during the irradiation and presented measurements.
The hit detection efficiency has been evaluated to~\qty{99.94+-0.05}{\percent} which is consistent with the efficiency of~\qty{99.96}{\percent} before irradiation~\cite{Bespin2024}.
\begin{figure}[ht]
  \centering
  \includegraphics[width=.9\columnwidth]{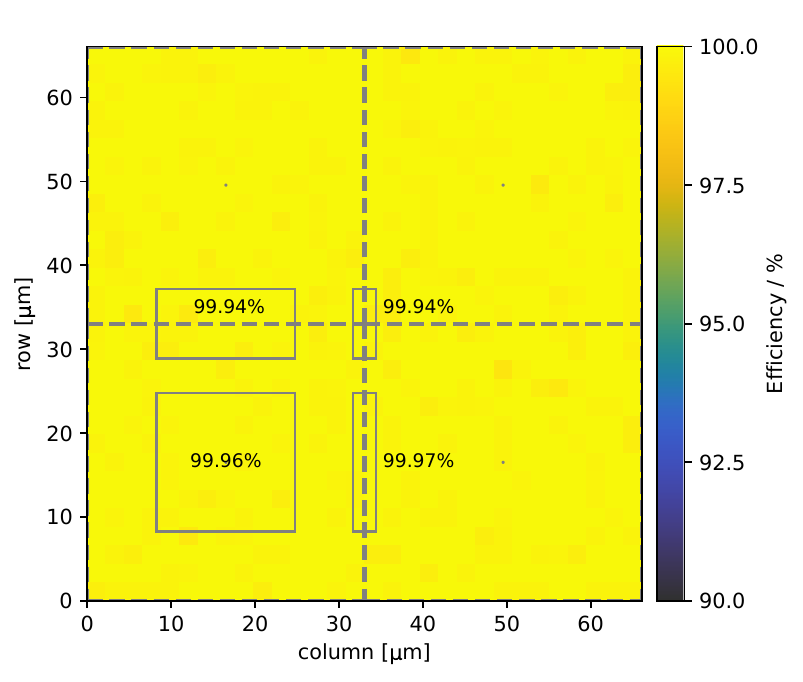}
  \caption{In-pixel efficiency map of TJ-Monopix2 after X-ray irradiation to~\qty{100}{\mega\rad}.}
  \label{fig:tj_eff}
\end{figure}
Figure~\ref{fig:tj_eff} shows the homogeneous in-pixel efficiency map after a total dose of~\qty{100}{\mega\rad} with individual values for different pixel regions.
In addition, the time resolution of about~\qty{1}{\nano\second} before irradiation is matched~\cite{Bespin2024a}.

\section{Conclusion}

Both DMAPS were successfully X-ray irradiated to~\qty{100}{\mega\rad} and remained fully functional at the highest dose level.
The power consumption follows the characteristic and, therefore, expected course for NMOS transistors of the respective feature size.
By utilizing the in-pixel threshold tuning capabilities and adjusting the front-end settings, the overall performance shows no significant degradation.
The threshold, threshold dispersion, and equivalent noise charge (ENC) before and after irradiation are summarized in Table~\ref{tab:summary}.
\begin{table}[ht]
  \centering
  \begin{tabular}{rSSSS}
    \toprule
     & \multicolumn{2}{c}{{LF-Monopix2}} & \multicolumn{2}{c}{{TJ-Monopix2}}\\
     & {\qty{0}{\mega\rad}} & {\qty{100}{\mega\rad}} & {\qty{0}{\mega\rad}} & {\qty{100}{\mega\rad}} \\\midrule
     Thr. / \unit{\e}       & 2055 & 1983 & 230 & 254 \\
     Thr. disp. / \unit{\e} &   91 &  108 &   5 &   5 \\
     ENC / \unit{\e}        &   92 &  112 &   6 &  13 \\\bottomrule  \end{tabular}
  \caption{Summary table of achieved performance of LF-Monopix2 and TJ-Monopix2 at~\qtylist{0;100}{\mega\rad} irradiation dose.}
  \label{tab:summary}
\end{table}

\section{Acknowledgments}

Parts of the measurements have been performed at the Test Beam Facility at DESY Hamburg (Germany), a member of the Helmholtz Association (HGF).
This project has received funding from the European Union's Horizon 2020 Research and Innovation programme under the Marie-Sklodowska-Curie GA No. 675587 (STREAM), under GA No. 654168 (AIDA-2020), and GA No. 101004761 (AIDAinnova).

\bibliographystyle{elsarticle-num-names} 
\bibliography{bibliography.bib}

\end{document}